# Simulations of magnetic and magnetoelastic properties of $Tb_2Ti_2O_7$ in paramagnetic phase

**V V Klekovkina[1], A R Zakirov[1], B Z Malkin[1] and L A Kasatkina[2]**

[1] Kazan Federal University, Kazan 420008, Russia
[2] Kazan State Technical University, Kazan 420111, Russia

E-mail: Vera.Klekovkina@gmail.com

**Abstract**. Magnetic and magnetoelastic properties of terbium titanate pyrochlore in paramagnetic phase are simulated. The magnetic field and temperature dependences of magnetization and forced magnetostriction in $Tb_2Ti_2O_7$ single crystals and polycrystalline samples are calculated in the framework of exchange charge model of crystal field theory and a mean field approximation. The set of electron-deformation coupling constants has been determined. Variations of elastic constants with temperature and applied magnetic field are discussed. Additional strong softening of the crystal lattice at liquid helium temperatures in the magnetic field directed along the rhombic symmetry axis is predicted.

## 1. Introduction

Rare earth titanates $R_2Ti_2O_7$ crystallizing in the pyrochlore lattice (the rare earth ions $R^{3+}$ form a sublattice of corner sharing regular tetrahedrons) are extensively studied during last ten years (for a review, see [1]). These compounds refer to the class of geometrically frustrated magnets and exhibit a wide variety of magnetic behaviors at low temperatures such as spin liquids (R=Tb), spin ices (R=Dy, Ho) or non-collinear long-range orders (R=Gd, Er) [2]. Thus $Tb_2Ti_2O_7$ is especially interesting since the origin of its spin-liquid ground state remains mysterious for many years.

The titled compound does not show any magnetic long-range order down to 50 mK and remains dynamic with short range correlations [3-4]. Theoretical investigation within a random phase approximation showed that it should undergo a transition into a magnetic long-range ordered state at temperature about 1.8 K [5]. However, magnetic properties of rare earth compounds can be strongly affected by magnetoelastic interactions. In particular, these interactions bring crystal lattice deformations (parastriction) in paramagnetic phase in presence of an external magnetic field. A cooperative Jahn-Teller phase transition (spontaneous lowering of crystal lattice symmetry due to interactions between the rare earth ions through fields of static and dynamic deformations [6]) can suppress the magnetic ordering.

Experimental investigations of parastriction and anomalous elastic properties of $Tb_2Ti_2O_7$ at low temperatures were performed over twenty years ago in [7-10] where a theoretical interpretation of the measured magnetic field and temperature dependences of the lattice constants and the Young's modulus was presented as well. At liquid helium temperatures and in the magnetic field of 5 T, the relative change in length along the applied field direction in polycrystalline and single crystal samples of $Tb_2Ti_2O_7$ reaches giant value of $0.5 \cdot 10^{-3}$ [8] that is at least two orders of magnitude larger than in typical paramagnets. The single ion mechanism of the parastriction [11] usually dominates in insulating rare earth compounds, and the field induced lattice strains depend on the crystal field energies of a rare earth ion, electron-deformation coupling constants and elastic constants of a lattice. The authors of [8, 10] determined correctly the ground state of the $Tb^{3+}$ ions and predicted a relatively small gap of 16 K between the ground and the first excited crystal field doublets. The results of measurements were reproduced satisfactorily in terms of a purely single-ion origin with making use of several phenomenological parameters of magnetoelastic coupling, and the temperature $T_c \square 0.1$ K of the cooperative Jahn-Teller structural transition was estimated from calculations [9, 10]. High-resolution neutron scattering data do not show any signs of symmetry breaking down to temperatures of 0.4-0.1 K [12].

Recent high-resolution x-ray diffraction measurements of $Tb_2Ti_2O_7$ single crystal revealed significant effects of the magnetoelastic coupling on properties of the crystal lattice at low temperatures in pulsed magnetic fields [13]. Essential softening of elastic constants at low temperatures was found from ultrasonic measurements [14, 15].

In the present work, our goal is to construct the model of magnetic and magnetoelastic interactions in $Tb_2Ti_2O_7$ that would allow us to describe satisfactorily the observed temperature and field dependences of the parastriction and elastic constants in this fascinating compound.

The paper is organized as follows. In section 2, we discuss the theoretical background and give values of model parameters. In section 3, the calculated bulk magnetization, individual $Tb^{3+}$ magnetic moments in external magnetic fields, the temperature and field dependences of the parastriction and elastic constants are compared with experimental data available from literature. Finally, conclusions are given in section 4.

## 2. Theoretical background and model parameters

2.1. Structural properties

The $Tb_2Ti_2O_7$ crystal lattice structure is isometric with space group $Fd3m$ ($O_h^7$, No. 227). The $Tb^{3+}$ and $Ti^{4+}$ ions occupy $16d$ (1/2, 1/2, 1/2) and $16c$ (0, 0, 0) sites, respectively, and the oxygen ions are in $8b$ (3/8, 3/8, 3/8) (we will designate it as O1) and $48f$ ($x$, 1/8, 1/8) (O2) Wyckoff positions. The lattice structure is completely described by the lattice constant $a = 1.01694$ nm and the parameter $x = 0.3287$ [16]. Geometrically, the $Tb^{3+}$ ion is in a strongly distorted cubic polyhedron formed by two O1 and six O2 ions (local point symmetry is trigonal $\bar{3}m$ ($D_{3d}$)).

Crystallographic Cartesian coordinates (in $a/8$ units) of the four nonequivalent positions of rare earth ions in the unit cell are: $\mathbf{r}_1 = (1,1,1)$, $\mathbf{r}_2 = (-1,-1,1)$, $\mathbf{r}_3 = (-1,1,-1)$, $\mathbf{r}_4 = (1,-1,-1)$. Below we use local coordinate systems for rare earth ions in the sublattices specified by vectors $\mathbf{r}_n$ connected to the crystallographic frame by the following rotation matrices, respectively:

$$R_1 = \begin{pmatrix} \frac{-1}{\sqrt{6}} & \frac{-1}{\sqrt{6}} & \sqrt{\frac{2}{3}} \\ \frac{1}{\sqrt{2}} & \frac{-1}{\sqrt{2}} & 0 \\ \frac{1}{\sqrt{3}} & \frac{1}{\sqrt{3}} & \frac{1}{\sqrt{3}} \end{pmatrix} \quad R_2 = \begin{pmatrix} \frac{1}{\sqrt{6}} & \frac{1}{\sqrt{6}} & \sqrt{\frac{2}{3}} \\ \frac{-1}{\sqrt{2}} & \frac{1}{\sqrt{2}} & 0 \\ \frac{-1}{\sqrt{3}} & \frac{-1}{\sqrt{3}} & \frac{1}{\sqrt{3}} \end{pmatrix} \quad R_3 = \begin{pmatrix} \frac{-1}{\sqrt{6}} & \frac{1}{\sqrt{6}} & \sqrt{\frac{2}{3}} \\ \frac{-1}{\sqrt{2}} & \frac{-1}{\sqrt{2}} & 0 \\ \frac{1}{\sqrt{3}} & \frac{-1}{\sqrt{3}} & \frac{1}{\sqrt{3}} \end{pmatrix} \quad R_4 = \begin{pmatrix} \frac{1}{\sqrt{6}} & \frac{-1}{\sqrt{6}} & \sqrt{\frac{2}{3}} \\ \frac{1}{\sqrt{2}} & \frac{1}{\sqrt{2}} & 0 \\ \frac{-1}{\sqrt{3}} & \frac{1}{\sqrt{3}} & \frac{1}{\sqrt{3}} \end{pmatrix} \quad (1)$$

We will mark all quantities in local coordinate systems by prime.

2.2. Crystal field parameters
Hamiltonian of a rare earth ion in the external magnetic field $\mathbf{B}$ can be presented in the form:
$$H = H_0 + H_{CF} + H_Z + H_{el-def} . \qquad (2)$$

Here $H_0$ is the Hamiltonian of a free $Tb^{3+}$ ion, operating in the total space of 3003 states of the electronic $4f^8$ configuration, $H_{CF}$ is the crystal field Hamiltonian, $H_{el\text{-}def}$ is the Hamiltonian of the electron-deformation interaction specified below, $H_Z = \mu_B \mathbf{B}_{eff}(\mathbf{L}+2\mathbf{S})$ is the electronic Zeeman energy, $\mu_B$ is the Bohr magneton, $\mathbf{L}$ and $\mathbf{S}$ are the electronic orbital and spin moments, respectively. Anisotropic exchange and magnetic dipolar interactions between the $Tb^{3+}$ ions were taken into account by making use of the mean field approximation. The effective magnetic field experienced by the $Tb^{3+}$ ions belonging to the $n$-sublattice ($n$=1-4) is written as [17]:

$$\mathbf{B}_{eff}(n) = \mathbf{B} - \mathbf{B}_d + \sum_{n'} \mathbf{Q}(n,n') \boldsymbol{\mu}(n')$$
$$+ \lambda_\| \sum_p (\mathbf{r}_{np}^- \cdot \boldsymbol{\mu}(p)) \mathbf{r}_{np}^- / (r_{np}^-)^2 + \lambda_{\perp 1} \sum_p (\mathbf{r}_{np}^+ \cdot \boldsymbol{\mu}(p)) \mathbf{r}_{np}^+ / (r_{np}^+)^2 + \lambda_{\perp 2} \sum_p (\mathbf{R}_{np} \cdot \boldsymbol{\mu}(p)) \mathbf{R}_{np} / (R_{np})^2. \quad (3)$$

Here $\boldsymbol{\mu}(n)$ is the thermal average of the magnetic moment $\mathbf{m} = -\mu_B(\mathbf{L}+2\mathbf{S})$ of the $Tb^{3+}$ ion with the basis vector $\mathbf{r}_n$ of the corresponding sublattice, $n$ and $p$ =1-6 number four magnetic ions in the unit cell and their nearest neighbors, respectively, $\mathbf{r}_{np}^\pm = \mathbf{r}_n \pm \mathbf{r}_p$, $\mathbf{R}_{np} = [\mathbf{r}_n \times \mathbf{r}_p]$. The second line in (3) corresponds to the exchange field defined by three independent parameters. Values of these parameters, $\lambda_\| = -0.079$ T/$\mu_B$, $\lambda_{\perp 1} = \lambda_{\perp 2} = -0.06$ T/$\mu_B$, were determined in [17] by fitting the simulated dc-susceptibility of $Tb_2Ti_2O_7$ to experimental data. Components of the tensor $\mathbf{Q}$ are the corresponding dipole lattice sums. In particular, the nonzero components of this tensor are $Q_{\alpha\alpha}(n,n) = 1$, $Q_{xx}(1,2) = Q_{yy}(1,2) = 2.0346$, $Q_{xy}(1,2) = 3.4522$, $Q_{zz}(1,2) = -1.0692$ in units of $4\pi/3v$ where $v = a^3/4$ is the unit cell volume (other components can be obtained using symmetry properties of the lattice). $\mathbf{B}_d$ is the demagnetizing field, in the present work we consider homogeneous demagnetizing fields collinear to the external field:

$$\mathbf{B}_d = \frac{4\pi N}{3vB^2} \sum_n (\boldsymbol{\mu}(n) \cdot \mathbf{B}) \mathbf{B}. \quad (4)$$

Here $N$ is the demagnetizing factor ($N$=1 for a sample of spherical shape).

The crystal field Hamiltonian $H_{CF}$ for rare earth ions in the pyrochlore lattice is determined by six crystal field parameters $B_p^k$. In the local Cartesian systems of coordinates with the $z'$-axes directed along the corresponding crystal ternary axes and the local $x'$-axes lying in the plane containing the crystal $z$-axis and the basis vector $\mathbf{r}_n$, the crystal field Hamiltonian $H_{CF}$ has the following form:

$$H_{CF} = B_2^0 O_2^0 + B_4^0 O_4^0 + B_6^0 O_6^0 + B_4^3 O_4^3 + B_6^3 O_6^3 + B_6^6 O_6^6 \quad (5)$$

where operators $O_p^k$ ($k \leq p$) are similar to Stevens' operators [18] and can be written as linear combinations of spherical tensor operators $C_k^{(p)}$ [19]. In particular, $O_p^0 = a_{p0} C_0^{(p)}$, $O_p^{2k-1} = -a_{p\,2k-1}(C_{2k-1}^{(p)} - C_{-2k+1}^{(p)})$, $O_p^{-2k+1} = i a_{p\,2k-1}(C_{2k-1}^{(p)} + C_{-2k+1}^{(p)})$, $O_p^{2k} = a_{p\,2k}(C_{2k}^{(p)} + C_{-2k}^{(p)})$, $O_p^{-2k} = -i a_{p\,2k}(C_{2k}^{(p)} - C_{-2k}^{(p)})$ for $k$=1, 2, 3; numerical factors equal $a_{20} = 2$; $a_{21} = 1/\sqrt{6}$; $a_{22} = 2/\sqrt{6}$; $a_{40} = 8$; $a_{41} = 2/\sqrt{5}$; $a_{42} = 4/\sqrt{10}$; $a_{43} = 2/\sqrt{35}$; $a_{44} = 8/\sqrt{70}$; $a_{60} = 16$; $a_{61} = 8/\sqrt{42}$; $a_{62} = 16/\sqrt{105}$; $a_{63} = 8/\sqrt{105}$; $a_{64} = 16/\sqrt{126}$; $a_{65} = 8/\sqrt{693}$; $a_{66} = 16/\sqrt{231}$.

The parameters $B_p^k$ of the crystal field were calculated as sums of Coulomb and exchange contributions in exchange charge approximation [19]:

$$B_p^k = B_{pq}^k + B_{pS}^k. \quad (6)$$

The Coulomb and exchange contributions are determined by the following formulae [19], respectively:

$$B^k_{pq} = -K^k_p \sum_{Ls} e^2 q_s (1-\sigma_p) \langle r^p \rangle O^k_p(\Theta_{Ls}\Phi_{Ls})/R_{Ls}^{p+1}, \qquad (7)$$

$$B^k_{pS} = K^k_p \sum_{v} \frac{2(2p+1)}{7} \frac{e^2}{R_v} S_p(R_v) O^k_p(\Theta_v \Phi_v), \qquad (8)$$

Here $K^0_p = (1/a_{p0})^2$, $K^k_p = \frac{1}{2}(1/a_{pk})^2$ for $k \neq 0$, $R_{Ls}$, $\Theta_{Ls}$, $\Phi_{Ls}$ are spherical coordinates of the ion belonging to the $s$-sublattice at the unit cell $L$, $R_v$, $\Theta_v$, $\Phi_v$ are the same for the ligand, $eq_s$ are the ion charges, $\sigma_p$ are the shielding factors, $\langle r^p \rangle$ are the moments of the 4$f$ electron space distribution, $e$ is the proton charge, $S_p(R_v)$ are bilinear forms constructed from the overlap integrals of 4$f$ wave functions and oxygen 2$s$-, 2$p\sigma$-, 2$p\pi$-functions with dimensionless phenomenological model parameters $G_s$, $G_\sigma$, $G_\pi$:

$$S_p(R_v) = G_s(S_s(R_v))^2 + G_\sigma(S_\sigma(R_v))^2 + [2 - p(p+1)/12] G_\pi(S_\pi(R_v))^2. \qquad (9)$$

**Table 1.** Crystal field parameters and energy levels $E$ of the Tb$^{3+}$ ground multiplet $^7F_6$.

| $p$ | $k$ | $B^k_p$ (cm$^{-1}$) | | | Symmetry | $E$ (cm$^{-1}$) | |
|---|---|---|---|---|---|---|---|
| | | This work | [17] | [20] | | Calculated | Measured [21] |
| 2 | 0 | 218.5 | 219 | 220 | $\Gamma_3$ | 0 | 0 |
| 4 | 0 | 320.6 | 319.4 | 317 | $\Gamma_3$ | 11.1 | 12.1 |
| 4 | 3 | -2188.7 | -2188.9 | -2174 | $\Gamma_2$ | 77.4 | 84.3 |
| 6 | 0 | 51.2 | 52.6 | 53 | $\Gamma_1$ | 121.1 | 118.8 |
| 6 | 3 | 777 | 807 | 807 | $\Gamma_3$ | 282.9 | |
| 6 | 6 | 809 | 779 | 807 | $\Gamma_2$ | 320.6 | |
| | | | | | $\Gamma_1$ | 323.9 | |
| | | | | | $\Gamma_3$ | 440.2 | |
| | | | | | $\Gamma_1$ | 512.7 | |

Numerical values of the overlap integrals and the shielding factor $\sigma_2$ were taken from [20]. Values of charges $eq_s$ and parameters $G_\alpha$ were chosen so that the crystal field parameters would be close to ones from [17, 20] that agree satisfactorily with the available information on crystal field energies of the Tb$^{3+}$ ions in Tb$_2$Ti$_2$O$_7$ [21]. The Coulomb contributions to the crystal field parameters were calculated using effective point charges 2.59 (Tb$^{3+}$), 3.39 (Ti$^{4+}$), -1.64 (O1$^{2-}$) and -1.72 (O2$^{2-}$) in units of the proton charge. The exchange charge contributions were calculated using values of the parameters $G_\alpha$ independent on the type of bonding $G_\sigma = G_s = G_\pi = 6.52$ for O1 ligands and $G_\sigma = G_s = G_\pi / 0.67 = 10.38$ for O2 ligands.

The obtained set of crystal field parameters is compared with the literature data in table 1. The single ion Hamiltonian $H_s = H_0 + H_{CF}$ was diagonalized numerically. The lowest 13 crystal field energies of the ground $^7F_6$ multiplet and the corresponding irreducible representations of the $D_{3d}$ point symmetry group are presented in table 1.

2.3. Electron-deformation interaction

The Hamiltonian of the linear electron-deformation interaction can be written as follows [19]:

$$H_{el-def} = \sum_{pk,\alpha\beta} B^k_{p,\alpha\beta} e'_{\alpha\beta} O^k_p = \sum_{pk,ijv} B^k_{p,v}(\Gamma^j_i) e'_v(\Gamma^j_i) O^k_p. \qquad (10)$$

Here $e'_{\alpha\beta}$ are components of the macroscopic strain tensor; $i$, $j$ and $\nu$ are type, number and row of an irreducible representation $\Gamma$ of the symmetry group $D_{3d}$, correspondingly. The symmetrized linear combinations of the components of the strain tensor are:

$$e'(\Gamma_1^1) = e'_{xx} + e'_{yy}; \quad e'(\Gamma_1^2) = e'_{zz}; \tag{11}$$

$$e'_1(\Gamma_3^1) = e'_{xx} - e'_{yy}; \quad e'_2(\Gamma_3^1) = 2e'_{xy}; \tag{12}$$

$$e'_1(\Gamma_3^2) = 2e'_{xz}; \quad e'_2(\Gamma_3^2) = -2e'_{yz}. \tag{13}$$

**Table 2.** Parameters of the electron–deformation interaction $B_{p,\lambda}^k(\Gamma_i^j)$ (cm$^{-1}$).

| pk | $B_p^k(\Gamma_1^1)$ | $B_p^k(\Gamma_1^2)$ | pk | $B_{p,1}^k(\Gamma_3^1)$** | $B_{p,1}^k(\Gamma_3^2)$** | pk | $B_{p,1}^k(\Gamma_3^1)$** | $B_{p,1}^k(\Gamma_3^2)$** |
|---|---|---|---|---|---|---|---|---|
| 2 0 | 2320 (3770)* | -6160 (-3960)* | 2 1 | 3950 (-1215)* | 6440 (5540)* | 2 2 | -2110 (-2910)* | 1976 (2480)* |
| 4 0 | -230 | -1860 | 4 1 | -2450 | 2585 | 4 2 | 665 | -1434 |
| 4 3 | 10050 | -2610 | 6 1 | 1228 | 1160 | 6 2 | -133 | 666 |
| 6 0 | 27 | -391 | 4 4 | -4810 | 1120 | 6 5 | 2174 | -138 |
| 6 3 | -3595 | 79 | 6 4 | 813 | -659 | | | |
| 6 6 | -3500 | -428 | | | | | | |

*- final values of coupling constants determined from the fitting procedure
**- $B_{p,2}^{-k}(\Gamma_3^j) = -B_{p,1}^k(\Gamma_3^j)$ for $k$=1, 4; $B_{p,2}^{-k}(\Gamma_3^j) = B_{p,1}^k(\Gamma_3^j)$ for $k$=2, 5

Parameters $B_{p,\alpha\beta}^k$ are related to the crystal field parameters by the following expression:

$$B_{p,\alpha\beta}^k = \frac{1}{2}\sum_{Ls}\left[X'_\alpha(Ls)\frac{\partial B_p^k}{\partial X'_\beta(Ls)} + X'_\beta(Ls)\frac{\partial B_p^k}{\partial X'_\alpha(Ls)}\right] \tag{14}$$

where $X'_\alpha(Ls)$ are the ion coordinates. It should be noted that we neglect additional terms in (10) caused by the interaction of the rare earth ions with sublattice displacements [19]. Nonzero electron-deformation parameters calculated according to (14) are presented in table 2.

Deformation of a crystal lattice in the magnetic field reduces free energy of a crystal and changes magnetic moments of rare earth ions. Parastriction in Tb$_2$Ti$_2$O$_7$ accept giant values [7, 8], and one may expect that the corresponding contributions into the magnetization are essential. The free energy per unit volume of the homogeneously deformed cubic lattice of Tb$_2$Ti$_2$O$_7$ equals

$$F_{lat} = \frac{1}{2}C_{\alpha\beta\gamma\delta}e_{\alpha\beta}e_{\gamma\delta} = \frac{1}{2}C(A_{1g})e(A_{1g})^2 + \frac{1}{2}C(E_g)\sum_{\lambda=1,2}e_\lambda(E_g)^2 + \frac{1}{2}C(F_{2g})\sum_{\lambda=1,2,3}e_\lambda(F_{2g})^2 \tag{15}$$

where linear combinations of the components of the strain tensor in the crystallographic frame

$$e(A_{1g}) = e_{xx} + e_{yy} + e_{zz}, \tag{16}$$

$$e_1(E_g) = e_{xx} - e_{yy}, \, e_2(E_g) = (2e_{zz} - e_{xx} - e_{yy})/\sqrt{3}, \tag{17}$$

$$e_1(F_{2g}) = e_{xy}, \, e_2(F_{2g}) = e_{yz}, \, e_3(F_{2g}) = e_{zx} \tag{18}$$

transform according to irreducible representations $A_{1g}$, $E_g$, $F_{2g}$ of the cubic symmetry group $O_h$, and $C_{\alpha\beta\gamma\delta}$ ($C_{\alpha\alpha\alpha\alpha}=C_{11}$, $C_{\alpha\alpha\beta\beta}=C_{12}$, $C_{\alpha\beta\alpha\beta}=C_{44}$) are the elastic constants which have been measured recently in [14, 15] in the temperature range from 1.5 to 300 K. The symmetrized elastic moduli $C(A_{1g})=(C_{11}+2C_{12})/3$, $C(E_g)=(C_{11}-C_{12})/2$ and $C(F_{2g})=4C_{44}$ refer to full symmetrical, tetragonal and trigonal deformations, respectively. For each rare earth $n$-sulattice, components of the strain tensor in the local and crystallographic frames are connected by the transformations

$$e'^{(n)}_{\alpha\beta} = \sum_{\gamma\delta} R_{n,\alpha\gamma} R_{n,\beta\delta} e_{\gamma\delta}. \tag{19}$$

Substitution of (19) into (10) allows us to obtain the free energy of the magnetic subsystem $F_M(\boldsymbol{B}, e_{\alpha\beta}) = -\frac{k_B T}{v} \sum_{n=1}^{4} \ln\{\text{Tr}\exp[-H_n(\boldsymbol{B}, e_{\alpha\beta})/k_B T]\}$ as a function of external magnetic field and macroscopic strains (here $k_B$ is the Boltzman constant, $H_n$ is the Hamiltonian (2) of the $Tb^{3+}$ ion belonging to the $n$-sublattice). From the minimum conditions of the total free energy of a crystal, $\partial(F_{lat} + F_M)/\partial e_{\alpha\beta} = 0$, and the definition of the magnetization $M_\alpha = \sum_n \mu_\alpha(n)/v = -\partial F_M/\partial B_\alpha$, we obtain the following system of self-consistent equations for strains and magnetic moments of the $Tb^{3+}$ ions induced by the external magnetic field:

$$e_\upsilon(\Gamma_i^j) = -\frac{1}{C(\Gamma_i^j)v} \sum_{n,pk,i'j'\upsilon'} B_{p,\upsilon'}^k(\Gamma_{i'}^{j'})\left(\langle O_p^k \rangle_n - \langle O_p^k \rangle_{n,\boldsymbol{B}=0}\right) T_{ij\upsilon}^{i'j'\upsilon'}(n), \tag{20}$$

$$\boldsymbol{\mu}(n) = -\mu_B \langle \boldsymbol{L} + 2\boldsymbol{S} \rangle_n = -\mu_B \frac{\text{Tr}[(\boldsymbol{L}+2\boldsymbol{S})\exp(-H_n(\boldsymbol{B},\{\boldsymbol{\mu},e(\Gamma)\})/k_B T)]}{\text{Tr}\exp[-H_n(\boldsymbol{B},\{\boldsymbol{\mu},e(\Gamma)\})/k_B T]}. \tag{21}$$

Here $T_{ij\upsilon}^{i'j'\upsilon'}(n) = \partial e'^{(n)}_{\upsilon'}(\Gamma_{i'}^{j'})/\partial e_\upsilon(\Gamma_i^j)$, $\langle ... \rangle_n$ and $\langle ... \rangle_{n,\boldsymbol{B}=0}$ are the thermal averages with the Hamiltonian of the $Tb^{3+}$ ion at the site $n$ in the magnetic field $\boldsymbol{B}$ and in zero field, respectively. The system of equations (20), (21) was solved using the method of successive approximations for each value of the field $\boldsymbol{B}$ and temperature $T$.

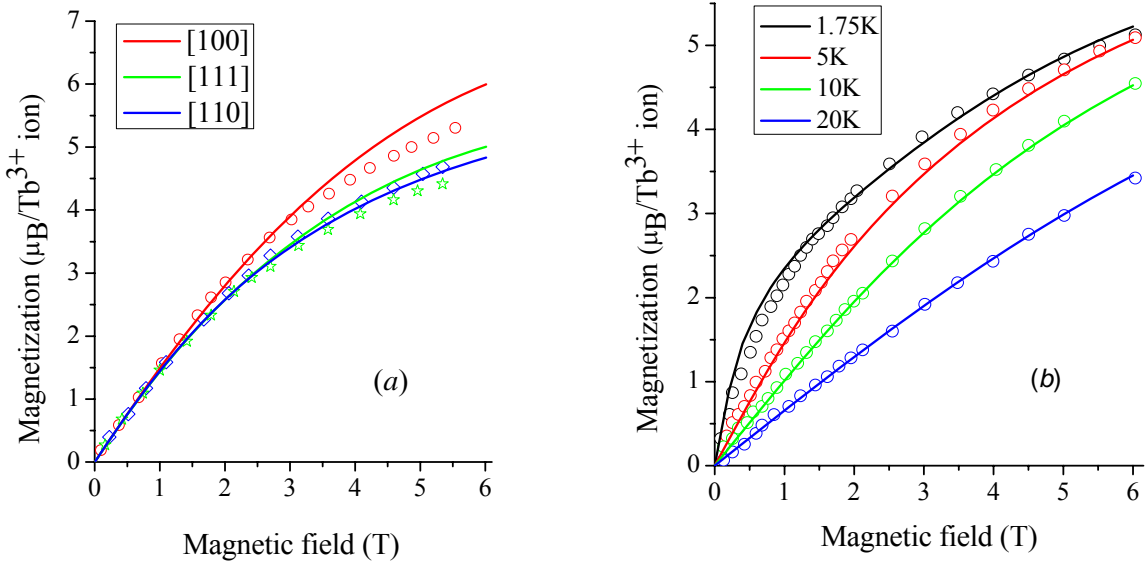

**Figure 1.** Measured (symbols) and calculated (solid curves) field dependences of the magnetization of $Tb_2Ti_2O_7$ single crystals for magnetic fields directed along the crystallographic axes at $T$=4.8 K (*a*) and of the magnetization of polycrystalline samples at different temperatures (*b*). Experimental points are digitized from [8] (*a*) and [25] (*b*).

2.4. Elastic constants
The contributions of the magnetic subsystem into the elastic constants were obtained from computations of internal stresses induced by small but finite deformations ($e_{0\upsilon}(\Gamma_i^j) \Box 10^{-6}$):

$$\Delta C(\Gamma_i^j) = \sum_n \frac{\sum_{pk,i'j'\upsilon'} B_{p,\upsilon'}^k(\Gamma_{i'}^{j'}) T_{ij\upsilon}^{i'j'\upsilon'}(n) \left[ \left\langle O_p^k \right\rangle_{n,e_0} - \left\langle O_p^k \right\rangle_{n,e_0=0} \right]}{e_{0\upsilon}(\Gamma_i^j) v}. \qquad (22)$$

Here $\langle...\rangle_{n,e_0}$ is the thermal average with the Hamiltonian $H_n$ containing the corresponding energy of electron-deformation interaction $\sum_{pk,i'j'\upsilon'} B_{p,\upsilon'}^k(\Gamma_{i'}^{j'}) T_{ij\upsilon}^{i'j'\upsilon'}(n) e_{0\upsilon}(\Gamma_i^j) O_p^k$. Note that we neglect second order terms in the electron-deformation interaction and intersite interactions via the phonon field.

2.5. Thermal expansion
Due to changes of populations of the crystal field sublevels of the ground multiplet with temperature, the $Tb^{3+}$ ions contribute into the thermal expansion of the crystal lattice. The corresponding difference between the lattice constant values at temperatures $T_1$ and $T_2$ equals $\delta a(T_1, T_2) = \Delta a(T_1) - \Delta a(T_2)$ where

$$\Delta a(T) = -\frac{4a}{9C(A_{1g})v} \sum_{pk} [2B_p^k(\Gamma_1^1) + B_p^k(\Gamma_1^2)] \frac{\mathrm{Tr}[O_p^k \exp(-H_s/k_B T)]}{\mathrm{Tr}[\exp(-H_s/k_B T)]}. \qquad (23)$$

2.6. Parastriction
Forced magnetostriction arises in a paramagnet due to redistribution of electronic density of paramagnetic ions in the external magnetic field that results in displacements of the ions in the local environment of each paramagnetic centre from their equilibrium positions, thus linear dimensions of the sample change.

Relative elongation (contraction) of uniformly strained crystal lattice along the unit vector **d** equals
$$\Delta l/l = \sum_{\alpha\beta} e_{\alpha\beta} d_\alpha d_\beta. \qquad (24)$$

In particular, we obtain from (24) the longitudinal parastriction (**B**∥**d**) of a cubic crystal in the following form:

$$(\Delta l/l)_\| = \frac{1}{3} e(A_{1g}) + e_1(E_g)(n_x^2 - n_y^2) + \frac{\sqrt{3}}{6} e_2(E_g)(3n_z^2 - 1) + \\ 2e_1(F_{2g}) n_x n_y + 2e_2(F_{2g}) n_y n_z + 2e_3(F_{2g}) n_x n_z \qquad (25)$$

where $n_\alpha$ are the directional cosines of the field **B** in the crystallographic frame. In general case, the transversal parastriction depends on the vector **d** direction in the plane normal to **B**, however, it is isotropic for magnetic fields directed along the $C_3$ and $C_4$ symmetry axes:

$$\mathbf{B}\|[001] \quad (\Delta l/l)_\| = \frac{1}{3} e(A_{1g}) + \frac{1}{\sqrt{3}} e_2(E_g), \quad (\Delta l/l)_\perp = \frac{1}{3} e(A_{1g}) - \frac{1}{2\sqrt{3}} e_2(E_g); \qquad (26)$$

$$\mathbf{B}\|[111] \quad (\Delta l/l)_\| = \frac{1}{3} e(A_{1g}) + 2e_1(F_{2g}), \quad (\Delta l/l)_\perp = \frac{1}{3} e(A_{1g}) - e_1(F_{2g}). \qquad (27)$$

The transversal parastriction of a single crystal in the magnetic field directed along the $C_2$ symmetry axis is anisotropic:

$$\mathbf{B}\|[110] \quad (\Delta l/l)_\| = \frac{1}{3} e(A_{1g}) - \frac{1}{2\sqrt{3}} e_2(E_g) + e_1(F_{2g}), \qquad (28)$$

$$(\Delta l/l)_{\perp 1} = (\Delta l/l)_{[0\ 0\ 1]} = \frac{1}{3} e(A_{1g}) + \frac{1}{\sqrt{3}} e_2(E_g), \qquad (29)$$

$$(\Delta l/l)_{\perp 2} = (\Delta l/l)_{[1\ -1\ 0]} = \frac{1}{3} e(A_{1g}) - \frac{1}{2\sqrt{3}} e_2(E_g) - e_1(F_{2g}). \qquad (30)$$

It is possible to present the strain tensor by power series in components of the magnetic field. In particular, $e(A_{1g}) = f(B,T) + b(T)(B_x^4 + B_y^4 + B_z^4) + ...$, thus the full symmetrical strain, similarly to low symmetry strains, depends not only on the magnitude of the field, but on its direction as well. For a powder sample, the expression (25) should be averaged over magnetic field directions.

## 3. Results of simulations and discussion

Due to a phenomenological character of the crystal field model, the obtained set of electron-deformation coupling constants can be used only for the initial estimations of the parastriction. Six parameters from the total set of thirty, $B_{2,\lambda}^k(\Gamma_i^j)$, which define changes of the quadrupolar component of the crystal field in the deformed crystal lattice, have been varied to reproduce available data on the parastriction in single crystals [8, 13]. The necessity of introducing large corrections of the calculated parameters $B_{2,\lambda}^k(\Gamma_i^j)$ most likely steams from the neglect of the sublattice displacements in the deformed crystal lattice. The final values of the varied coupling constants (see table 2) depend essentially on the values of the elastic moduli used in the simulations. The measured field and temperature dependences of the elastic constants of $Tb_2Ti_2O_7$ presented in [14] and [15] are qualitatively the same, but numerical values are quite different. The results of ultrasonic measurements in [14] do not agree with the previously measured bulk modulus $B_0$ in the x-ray synchrotron diffraction studies [22] of the crystal structure under pressure (values of $(C_{11}+2C_{12})/3$ in [14] are about three times less than $B_0$ in [22]). Thus we have used in the present work the elastic constants measured in [15].

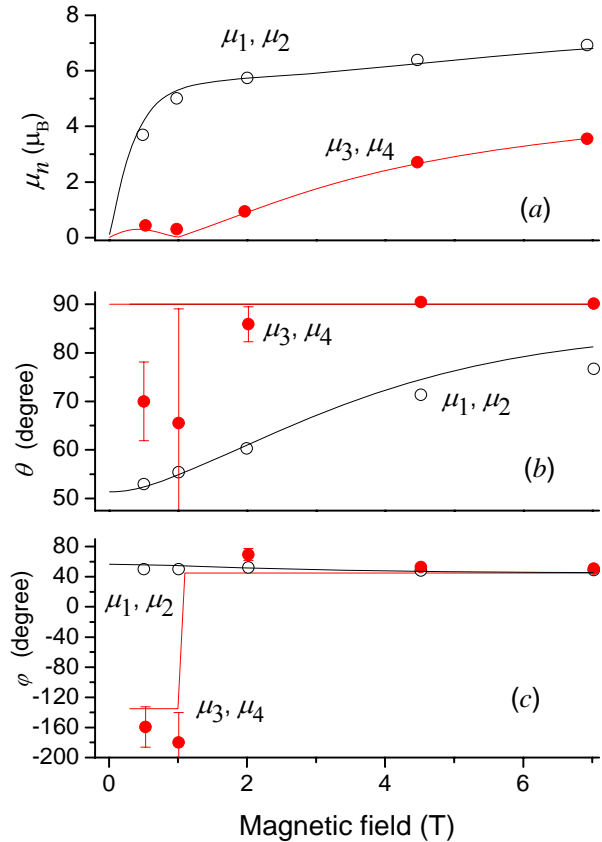

**Figure 2.** Field dependences of the $Tb^{3+}$ magnetic moments for magnetic field **B** ∥ [110] at temperature 1.6 K: calculated (solid curves) magnetic moments (*a*), angles $\theta$ (*b*) and $\varphi$ (*c*) between a moment and the [001] and [100] axes, respectively. Experimental data (symbols) are digitized from [26].

We calculated dependences of the magnetization and individual terbium magnetic moments on the external field and temperature to test our model. In arbitrary directed magnetic field, the $Tb^{3+}$ ions occupy four nonequivalent sites. Four terbium sublattices are equivalent for magnetic field parallel to a tetragonal axis of the lattice, but there are two nonequivalent sites with different magnetic moments if the external field is parallel to trigonal or rhombic axes. Figure 1(*a*) shows calculated field dependences of the magnetization of a single crystal for magnetic fields directed along $C_4$, $C_3$ and $C_2$ symmetry axes at temperature 4.8 K. From a comparison of the results of calculations with experimental data, it follows that neglect of magnetic interactions between the terbium ions brings essential overestimation of the calculated magnetization. Accounting for dipole-dipole and exchange interactions, we are able to describe satisfactorily the magnetization behavior in fields up to 4 T. In high magnetic fields directed along the $C_4$ or $C_3$ axes, the calculated magnetization is overestimated. Simulations were carried out for samples of cylindrical shape with the demagnetizing factor $N$=0.5. Note that the exchange interaction plays dominant role in formation of local magnetic fields while the dipole-dipole interaction brings relatively small contributions. In agreement with the experimental data [8], simulations reveal a small anisotropy of the magnetization of single crystals with "easy" direction being [100]. However, contrary to the experimental data [8, 23], the difference between calculated magnetizations $M(B,T)_{B\|[111]} - M(B,T)_{B\|[110]}$ is positive. The sign of this difference is stable against possible variations of the parameters of the exchange interaction between the nearest neighbor $Tb^{3+}$ ions consistent with the temperature dependences of the isotropic bulk and anisotropic site susceptibilities in $Tb_2Ti_2O_7$ [17].

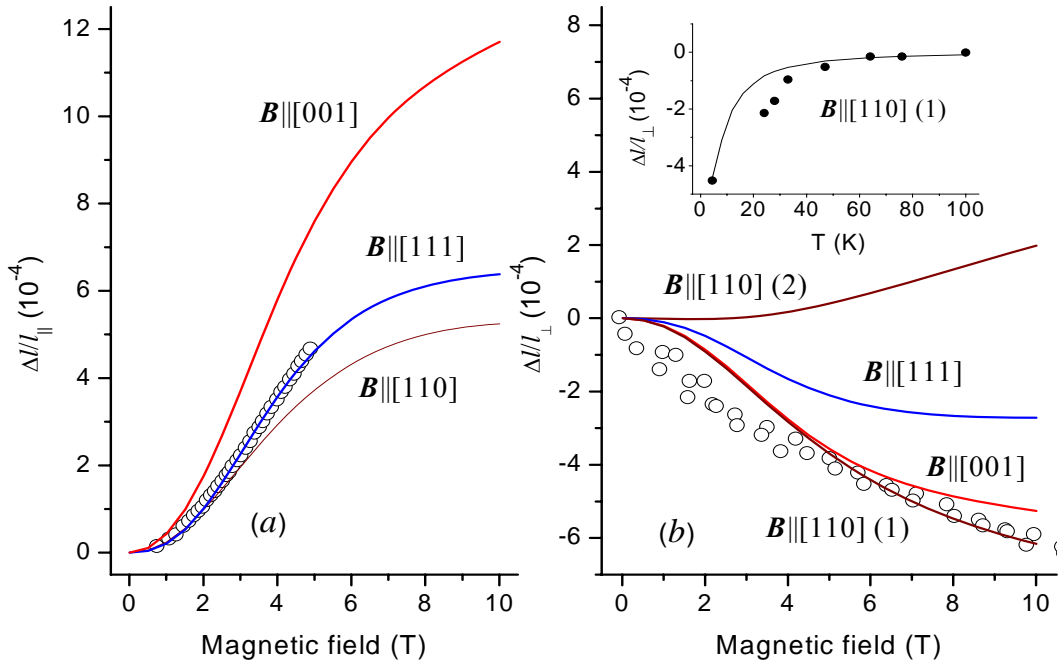

**Figure 3.** Calculated field dependences of the longitudinal (*a*) and transversal (*b*) parastriction of single crystals at temperature 4.2 K for $B\|[111]$ and 4.5 K for $B\|[001]$ and $B\|[110]$. Experimental data from [8] (*a*) and [13] (*b*) are represented by symbols. Inset: Calculated and measured [13] temperature dependences of the transversal parastriction $\Delta l / l_{[001]}$ in the field $B\|[110]$, $B$ = 6 T.

When varying independently the parameters of the effective field (3), we have obtained $\lambda_\| $ = -0.092 T/$\mu_B$, $\lambda_{\perp 1}$ = -0.058 T/$\mu_B$, $\lambda_{\perp 2}$ = -0.01 T/$\mu_B$; these values have been used in all simulations which we discuss here. Accounting for the allowed by symmetry contributions into the effective magnetic fields due to Dzyaloshinskii-Moriya interaction [24] does not help to solve the problem with the sign of

$M(B,T)_{B\|[111]} - M(B,T)_{B\|[110]}$. The absolute values of these differences are comparable to the positive contributions into the magnetization due to the magnetoelastic interaction (the inverse parastriction effect).

Figure 1(*b*) shows the field dependences of the magnetization (averaged over a sphere) of the polycrystalline sample at different temperatures. The calculated magnetization curves nearly match experimental data [25] down to temperature of 1.75 K.

Relative values of magnetic moments of the Tb$^{3+}$ ions at nonequivalent sites are very sensitive to the exchange coupling constants. Calculated field dependences of magnetic moments for magnetic field along the rhombic [110] axis at temperature of 1.6 K are shown in figure 2 (*a*). Figures 2 (*b*) and 2 (*c*) show the angles $\theta$ and $\varphi$ between the moments and the [001] and [100] axes of the cubic unit cell, respectively. Moments at $\alpha$-sites (*n*=1, 2; the magnetic field is declined from the local trigonal axis by the angle of ~35.3 degrees) relatively quickly saturate. As it is clear from the $\theta(B)$ dependence, as the magnetic field increases, $\alpha$-moments slowly rotate from their local easy axis toward the field direction [110] lying in the (001) plane, and the component of the $\alpha$-site moment along [001] becomes smaller. In agreement with experimental data [26], in the magnetic field ***B***||[110], the moments at $\beta$-sites (*n*=3, 4; the field is perpendicular to the local trigonal axis) undergo a "spin-melting" at *B* = 1.1 T where their magnitude vanishes. Moreover, the $\beta$-moments flip from a direction opposite to magnetic field to the direction parallel to field (see figure 2 (*c*)) at the same critical field. Note that simulations without taking into account the exchange interactions predict monotonous increasing of the $\beta$-moments with increasing magnetic field.

Figure 3 shows calculated field and temperature dependences of the forced magnetostriction of single crystals. It is seen that parastriction of single crystals depends strongly on a direction of the external magnetic field. Similarly to the magnetization, the longitudinal parastriction has the largest values in magnetic fields directed along the tetragonal axes. At liquid helium temperatures the transversal parastriction is negative and about two times less in magnitude than the longitudinal parastriction in the fields along the tetragonal and trigonal axes in agreement with the measurements in powder samples [8]. We have found very large anisotropy of the transversal parastriction for the magnetic field directed along the $C_2$ axis (in particular, [110], see figure 3 (*b*)). In this case, the transversal parastriction has extreme values along the $C_2$ axis ([1-10]) normal to the field (maximum, positive value) and along the $C_4$ axis ([001], minimum, negative value). Calculated temperature dependences of the parastriction (not shown) are also different for different directions of the applied field. With increasing temperature and correspondingly increasing population of the first excited doublet state of the Tb$^{3+}$ ions, the longitudinal parastriction in magnetic field applied along the [111] axis quickly decreases and changes its sign at $T \approx 16$ K. With further increasing temperature, it remains negative, but the absolute value slowly deceases. However, the longitudinal parastriction in magnetic fields parallel to $C_4$ and $C_2$ axes remains positive at all temperatures. The calculated temperature dependence of the transversal parastriction in the field ***B***||[110] agrees with experimental data [13] (see Inset in figure 3 (*b*)), but we obtained significantly overestimated absolute values of the transversal parastriction for ***B***||[001] which, according to [13], should be about two times less than for ***B***||[110].

Volume parastriction $\Delta V / V = e(A_{1g})$ is determined, similarly to the contributions into the lattice thermal expansion and the elastic bulk modulus $C(A_{1g})$, by parameters $2B_p^k(\Gamma_1^1) + B_p^k(\Gamma_1^2)$. Using values of these parameters presented in table 2, we obtained positive volume parastriction at all temperatures in agreement with the results of measurements in powder samples [8]. The calculated crystal lattice expansion $\delta a(24 \text{ K}, 4.2 \text{ K})/a \approx 0.45 \cdot 10^{-4}$ caused by the interaction between the lattice and the terbium subsystem does not agree with the anomalous negative thermal expansion below 20 K which has been reported for a crystal sample [27] but is consistent with the just published results of x-ray powder diffraction experiments [28]. Despite the correctly estimated thermal expansion, we obtained rather small contributions into the bulk modulus (absolute values are less than one GPa as compared with difference of ~22 GPa between the bulk moduli at 50 and 5 K measured in [15]). Note that the absolute value of the negative contribution into the bulk modulus due to linear electron-deformation interaction should have a maximum at $T \approx 12$ K (the full symmetrical local deformations $e'(\Gamma_1^j)$ mix two lower doublets in the energy spectrum of the $Tb^{3+}$ ions but do not split them), however, the corresponding minimum in the temperature dependence of $C(A_{1g})$ has not been revealed in [14] and [15].

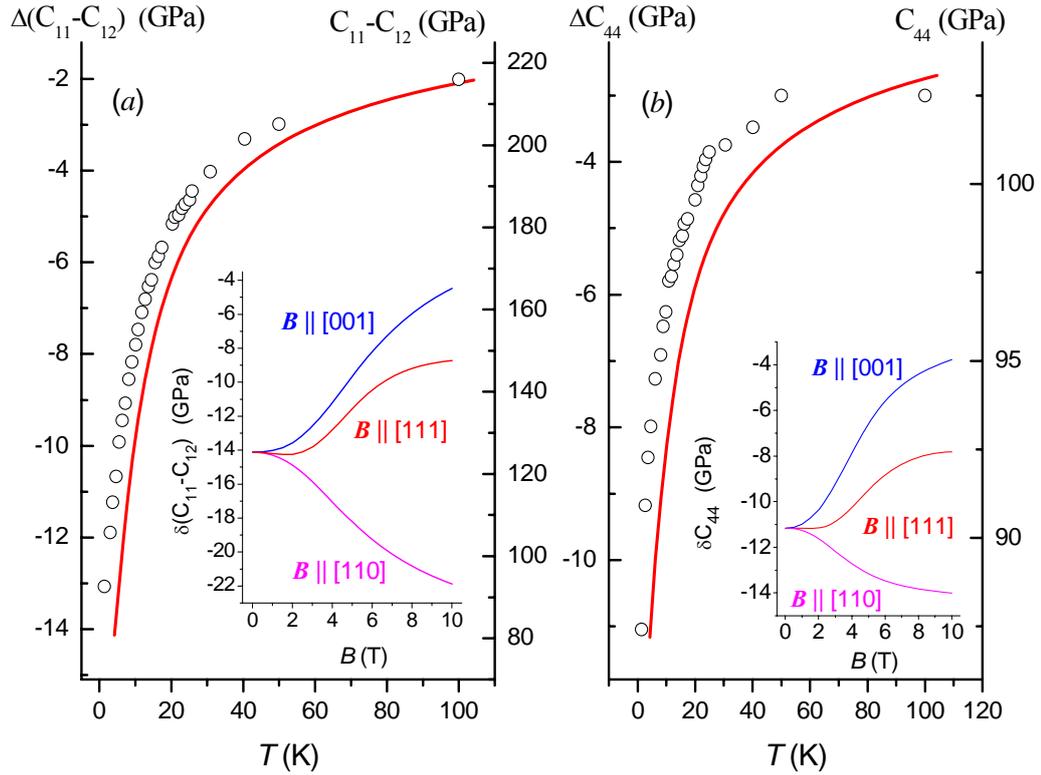

**Figure 4**. Temperature dependences of elastic constants measured in [15] (symbols) and calculated contributions into the elastic constants caused by linear electron-deformation interaction. Insets: calculated changes of elastic constants at temperature 4.2 K in magnetic fields directed along the $C_4$, $C_3$ and $C_2$ symmetry axes.

Temperature dependences of the elastic moduli $C(E_g)$ and $C(F_{2g})$ measured in [15] as well as the calculated contributions to these moduli induced by the electron-deformation interaction (10) are shown in figure 4. The calculated change of $C(E_g)$ in the temperature range from 100 down to 4.2 K is smaller than the measured one by an order of magnitude, but calculated and measured changes of the $C(F_{2g})$ modulus in the same temperature range have comparable values. Thus, it is possible that renormalization of the elastic multipolar susceptibilities of the $Tb^{3+}$ ions due to intersite interactions through the dynamic lattice deformations and second order terms in the electron-deformation interaction play essential role in responses of the magnetic subsystem on the lattice strains. Simulations of the changes of the elastic moduli in external magnetic fields, $\delta C_{ij}(\boldsymbol{B}) = C_{ij}(T,\boldsymbol{B}) - C_{ij}(T,\boldsymbol{B}=0)$, at low temperatures have allowed us to reveal rather large anisotropy of magnetoelastic effects (see Insets in figure 4): in agreement with the data presented in

[14], the magnetic field directed along the $C_4$ axis suppresses the lattice softening caused by the electron-deformation interaction, however, the field directed along the $C_2$ axis induces additional remarkable decrease of the elastic moduli $C(E_g)$ and $C(F_{2g})$ and may give rise to structural phase transition.

## 4. Conclusion

In the present work, a model has been derived to describe spectral, magnetic and magnetoelastic properties of terbium titanate pyrochlore single crystals (we have not discussed in details the forced magnetostriction in powder samples because of possible porosity). The model operates with three phenomenological parameters of the anisotropic exchange interaction between the nearest neighbor terbium ions, six crystal field parameters and thirty parameters which determine modulation of the crystal field by lattice strains. The parameters of the electron-lattice interaction have been estimated in the framework of the semi-phenomenological exchange charge model of the crystal field in ionic crystals, but six parameters of the quadrupolar component of the modulated crystal field have been varied to fit the experimental data on single crystal deformations in the magnetic fields available from literature. The fitting was performed by making use of the elastic constants of $Tb_2Ti_2O_7$ determined recently from ultrasonic measurements [15].

We reproduced satisfactorily main features of the temperature and field dependences of the magnetization and parastriction in $Tb_2Ti_2O_7$ which had been published earlier. Thus we have good reasons to believe that the giant anisotropy of the transversal parastriction and softening of the crystal lattice in magnetic fields parallel to the rhombic symmetry axes ([110], in particular) revealed from our simulations are real properties of $Tb_2Ti_2O_7$. However, some problems concerning more exact quantitative description of the magnetization, parastriction and changes of elastic constants with temperature remain unsolved. Additional studies of effects due to mixing of sublattice displacements corresponding to Raman active $A_{1g}$, $E_g$ and four $F_{2g}$ lattice vibrations [29] with macroscopic strains, intersite multipolar interactions and two-particle terms in the electron-deformation interaction are necessary, but these tasks are out of scopes of the present work.

The authors acknowledge support by the RFBR grant №09-02-00930 and by the Kazan Federal University in the frameworks of the project F11-22.